\documentclass[twocolumn,showpacs]{revtex4}

\usepackage{amsmath}
\usepackage{amsfonts}

\begin{document}

\title{Uncertainty relations describing the short-term violations of Lorentz
invariance: superluminal phenomena, particles transformations}
\author{Mark E.Perel'man}
\affiliation{\textit{Racah Institute of Physics, Hebrew University, Jerusalem, Israel }}

\keywords{Keywords: \textmd{uncertainty, Lorentz violation, superluminal, K$^{0}%
$-mesons, neutrino}}
\pacs{PACS numbers: 03.65.-w; 03.30.+p; 03.65.Xp; 13.20.Eb; 14.60.Lm}

\begin{abstract}
The refinement and specifications of time-energy uncertainty
relations have shown that the experimentally observed phenomena of
superluminal signaling are describable by such their form: $\Delta
E\Delta\tau\geq\pi\hbar$, where both standard deviations are
negative. When $\Delta\tau<0$, these evanescent photons would be
instantly tunneling from one light cone into another on the distance
$c\left\vert \Delta\tau\right\vert $. (This assertion, previously
proved via dispersion relations, is described here by the temporal
parameters of process.) Special forms of these relations describe
the transmutation of particles into their partners of bigger mass,
in $K^{0}$ and $B^{0}$ cases and at the $\nu$'s transmutations.
Thus, the violations of relativistic causality by evanescent
particles can be considered as the tunneling, as an analog of the
short-term violation of conservation laws at virtual transitions.
The absence of Lorentz invariance at such transitions can, probably,
allow violations of some other symmetries.

\end{abstract}

\date{\today}
\maketitle

\section{INTRODUCTION}

In the last years some possibilities of violations of the Lorentz invariance
and relativistic causality are actively discussing. On the one hand there are
the numerous experimental observations of superluminal, faster-than-$c$,
signal transfer at low energies (e.g. the reviews [1]). On the other hand the
discussions of such possibilities in the high energy physics are conducted,
they are connected with $CP$ or even $CPT$ violations and are usually
considered via addition of new terms in the standard Lagrangian of field
theories (e.g. [2]) and so on.

As it is proved in [3] the experimental data of superluminal transfer are
restricted to the condition: \textit{Superluminal transfer of excitations
(jumps) through the linear passive substance can affected by nothing, but the
instantaneous tunneling of virtual particles; distance of tunneling is of
order of half wavelength corresponding to energy deficit relative to the
nearest stable (resonance) state}.

This condition can be expressed as

$(\Delta E\Delta\tau)_{stable}\geq\pi\hbar$, \qquad(1.1)

where both standard deviations $\Delta E$ and $\Delta\tau$ must be
simultaneously negational and which formally looks like an extraordinary
special form of the uncertainty relations. The negativity of $\Delta\tau$ can
be interpreted as an advancing emission or as the instantaneous transferring,
i.e. the tunneling through classically forbidden space-time regions, from one
light cone into another.

Thus some questions should be considered: what sense can have such seemingly
violations of the general principle of locality and relativistic causality of
the theory, and thereafter: is the form (1.1) universal and so can it predict
analogical peculiarities in another phenomena, or such forms must be
specialized for different processes if they exist?

What can we say, firstly, if the apparent Lorentz violations in low and high
energy ranges exist, but they can be expressed in the form or via some analogs
of uncertainty principle?

Let's remember that although the virtual transitions violate the conservation
laws, short-termly or on the distances of tunneling, in the scope of
uncertainty principles, it does not present difficulties and does not provoke
discussions: it is well known that such violations do not lead to
possibilities of perpetuum mobiles construction, etc. And if analogously the
deviations from the Lorentz invariance would be restricted by some form of
quantum uncertainties or indeterminacies, it would only mean that along with
virtual states or particles the evanescent states or particles can be
manifested, but it does not mean any possibility of time arrow transformation
(the common or primitive causality). Such phenomenon would demonstrate only
the existence of one more quantum peculiarity of Nature, the specific
possibilities of quantum tunneling between light cones and so on, and does not
require any revision of the relativity's principles.

For an answer to the second question it must be taken into account that the
energy-time uncertainty principle is of more complicate nature than others.
The Heisenberg canonical form $\Delta A\cdot\Delta B\geq\hbar/2$, which is
obviously cited as the general form of uncertainty relation, is the
oversimplified one and many years ago was generalized on superconcept
relations by Robertson [4] and Schr\"{o}dinger [5]. It seems that the most
constructive form of such generalization was expressed by Mandelstam and Tamm
[6]. They had shown, in particular, that for decay processes the energy-time
uncertainties must be expressed as

$(\Delta E\Delta t)_{decay}\geq\pi\hbar/4$. \qquad(1.2)

In the recent years the attention to such forms was attracted by
investigations of so named squeezed light phenomena (e.g. [7] and references therein).

All it means the necessity of reconsideration of these uncertainty relations.

As the relation (1.1) is of a new form and can be directly tested by
experiments, we shall begin with its outline in the Section 2. Its deduction
in [3] for processes of superluminal transfer of excitations requires the
complicate dispersion relations, which can overshadow its direct sense.
Therefore its overview follows the temporal functions describing the time
delay at scattering process and the duration of particle formation (dressing),
i.e. all consideration goes via the kinematics of transferring. The deduction
in [3] was executed for photons only, and for widening such approach the
temporal functions for massive particles needed for further consideration are
written out in the Section 3.

The general consideration of uncertainty relations in the Section 4 confirms
both forms, (1.1) and (1.2), for corresponding processes. Along with them it
leads to more customary, but more specificated relation

$(\Delta E\Delta t)_{transmut}\geq\hbar/2$ \qquad(1.3)

for the processes of particles transmutation. The application of this relation
to processes of neutral mesons and neutrinos transmutations is briefly
considered in the Section 5. This examination would demonstrate that the
transmutations could be interpreted as the transitions into higher energy
(mass) state only, and just therefore, for example, the transitions of muonic
neutrino into electronic one can be absent or be suppressed. This relation
leads also to a new estimation of the neutrino mass.

These results and certain further perspectives are summed in the Conclusions.

\qquad

\section{TEMPORAL FUNCTIONS FOR MASSLESS PARTICLES}

The description of "superluminal" processes can be performed via two
analytically connected temporal functions.

The time duration of particle (photon) delay in the course of elastic
scattering is determined by the Wigner-Smith expression via the logarithmic
derivative of corresponding matrix element (e.g. [8]):

$\tau_{1}=$ Re$(\partial/i\partial\omega)\ln S(\omega,\mathbf{k})$. \qquad(2.1)

This magnitude is related to the scatterer and describes its state only.
(There are many different definition of delay duration, e.g. the reviews [9],
but for our discussion the general definition (2.1) seems enough.) Along with
this temporal magnitude the duration of gradual formation (the "dressing") of
formed particle should be determined. Such problem was initially considered,
in the very general form, by Bohr and Rosenfeld [10]. In the connection with
experiments such notion had been introduced for the first time by I. Frank
[11] in the semiclassical theory of \^{C}erenkov radiation: without it the
emission of discrete quanta at the uniform motion of charge, even with the
superluminal velocity in medium, was absolutely non-understood. By present
time some analogical problems compose the special direction in the radiation
and scattering theory (e.g. the review [12]).

For our purposes the duration of formation can be expressed in the form
similar (2.1) (its substantiation and more detailed statements are given in [13]):

$\tau_{2}=\operatorname{Im}(\partial/i\partial\omega)\ln S(\omega,\mathbf{k}%
)$. \qquad(2.2)

The joining of these two expressions leads to the equation:

$\frac{\partial}{i\partial\omega}S(\omega,\mathbf{k})=\tau S(\omega
,\mathbf{k})$, \qquad(2.3)

which can be considered as the analog of the Schr\"{o}dinger equation for
$S$-matrix of interaction with a temporal operator $\tau$ that plays the role
of Hamiltonian.

The consideration of temporal functions of QED would begin from the temporal
features of the simplest photon causal propagator (the Feynman gauge,
$\eta\rightarrow0+$):

$D_{c}(\omega,\mathbf{k})=4\pi/(\omega^{2}-k^{2}+i\eta)$, \qquad(2.4)

which leads to the expressions for durations of delay and of formation:

$\tau_{1}=-2\pi\omega\ \delta(\omega^{2}-\mathbf{k}^{2})\rightarrow-\pi
\delta(\omega-|\mathbf{k}|)$, \qquad(2.5)

$\tau_{2}=2\omega/(\omega^{2}-\mathbf{k}^{2})\rightarrow1/(\omega
-|\mathbf{k}|)$. \qquad(2.5')

As the propagator (2.4) does not include parameters of scatterer, the function
$\tau_{1}$ descriptively shows that photon can be absorbed or emitted only
completely; already this form contradicts possibility of its gradual
evolution. The function $\tau_{2}$ qualitatively corresponds to the
uncertainty principle, is twice bigger its usual form and shows the
possibility of retarded, at $\omega>|\mathbf{k}|$, or advanced, at
$\omega<|\mathbf{k}|$, emissions of photon; just the advanced form must be
interpreted as the instantaneous jump onto corresponding distance. (Remember
in this connection that the causal propagators overstep the limits of cone.)

These definitions are of the general character. Let's concretize them for
superluminal processes by considering the $(\omega,\mathbf{r})$%
-representation, from which would be deduced the distances of "superluminal"
transitions. It must be noted only that the temporal functions $\tau
(\omega,\mathbf{r})$ and $\tau(\omega,\mathbf{k})$ are not simply connected by
the Fourier transformation; their interrelation is partly considered in [13],
but here we shall use both for the qualitative analysis.

The causal propagator of QED $D_{c}(t,\mathbf{r})=\overline{D}(t,\mathbf{r}%
)+iD_{1}(t,\mathbf{r})$, where the first Green function is supported in the
light cone, but the second one oversteps its limits and therefore is of the
prime interest for us. In the mixed representation $D_{1}(\omega
,\mathbf{r})=(1/2\pi)\sin(|\omega|r)$ and the corresponding temporal function:

$\tau(\omega,\mathbf{r})=(\partial/i\partial\omega)\ln D_{1}(\omega
,\mathbf{r})=-ir\cot(\omega r)$ \qquad(2.6)

or

$\tau_{1}(\omega,\mathbf{r})=0$;\qquad\ $\tau_{2}(\omega,\mathbf{r})$
$=-r\cot(\omega r)$. \qquad(2.6')

These expressions implicitly show, that such process can go without delay and
that under definite values of $\omega r$ the duration of formation can be
negational, i.e. along with the "normal" transitions (2.6) can describes the
instantaneous jumps of transferred excitation.

The Coulomb field infinitely, as the static one, is in the "undressed" state,
and therefore the subtraction of the Coulomb pole $1/\omega$ in (2.6') is
needed. This subtraction can be performed with the decomposition of cotangent:

$\cot(x)=1/x+\sum_{1}^{\infty}(2x)/(x^{2}-\pi^{2}n^{2})$.

It leads to the renormalized expression for $\tau_{2}$:

$\tau_{2}^{(renorm)}(\omega,\mathbf{r})=-\sum_{1}^{\infty}2\omega
r/(\omega^{2}r^{2}-\pi^{2}n^{2})$, \qquad(2.7)

which shows that the first pole of (2.7) is at the point $\omega r=\pi$ (near
to resonance the substitution $\omega\rightarrow\Delta\omega$ can be made).
From (2.7) follows the (minimal) formation path for photon:

$\Delta\ell\simeq\pi c/\left\vert \Delta\omega\right\vert $. \qquad(2.8)

As this process is instantaneous, it corresponds to the jump of photon at the
act of formation on the distance $\pi c/|\Delta\omega|$ or $\lambda/2$, if
$\Delta\omega\rightarrow\omega$. Thus, the expressions (2.5-8) visually
outline the main part of the theorem [3] cited above, at any rate for
nonresonant case.

Let us consider the properties of media, in which the manifestation of
superluminal phenomena can be possible at definite frequencies. Note here that
in accord with (2.7), the frequencies domains of superluminal and subluminal,
including so named "slow light", are very close, it illustrates the
experimental difficulties with searching of "superluminal ranges".

As the instantaneous transferring is possible for virtual particles only,
their formation length must be not lesser the free path length of photon
$\ell$=1/$\rho\sigma$, where $\rho$ is the density of scatterers (free and
valent electrons), $\sigma$ is the total cross-section of single $\gamma$-$e$
scattering. This indispensable condition presents the possibility of virtual
excitations exchange between scatterers, i.e. it is the necessary condition of tunneling:

$\pi c/|\Delta\omega|\ >1/\rho\sigma$, \qquad(2.9)

which should be further specified for concrete cases.

At low, optical frequencies, sufficiently remote from any resonance, the
unique process, at which can take place the reemission with needed subsequent
dressing, is the Compton scattering described by the Thompson cross-section
$\sigma_{T}$. Hence at such conditions, when $\Delta\omega\rightarrow\omega$,
the superluminal phenomena are possible for sufficiently long wavelengths only:

$\lambda>2/\rho\sigma_{T}$. \qquad(2.10)

It shows that for condensed substances with $\rho~\sim10^{22}\div10^{23}$
these processes can go at frequencies of order of hundreds MHz or lower
(notice that the first regular investigations of superluminal phenomena have
been carried out just in the microwave region [14]).

Close to resonance, when $\sigma_{res}\simeq\lambda^{2}\Gamma^{2}/\pi
\lbrack\Delta\omega^{2}+\Gamma^{2}/4]$ (factors connected with angular moments
are omitted), the conditions of observability of superluminal signals become
easier. For $(\omega-|k|)\simeq(\omega-\omega_{0})\equiv-\left\vert
\Delta\omega\right\vert $ below the resonance and at $|\Delta\omega|<\Gamma$
the condition (2.1) takes the form

$|\Delta\omega|\ \leq c\rho\lambda^{2}$ \qquad(2.11)

and suggests the possibility of these phenomena in the wide range of
scatterers density, from rare gases till condensed states.

At $|\Delta\omega|\ >>\Gamma$ we return essentially to relations close to (2.10)

Let's consider now the possibility to describe these properties in the frame
of scattering theory via the group index of refraction, i.e. as the
macroscopic features [15].

The time intervals needed for photons' passing through distances $L$ in vacuum
and in medium are, respectively,

$T=L/c$; \qquad$T_{1}=L/c+N\tau_{1}=T+\Delta T$, \qquad(2.12)

where $N=L/\ell$ is the mean number of elastic scattering of photons on their
path length, $\ell=1/\rho\sigma$ is the free path length of photons. Therefore
the group index of refraction is given by the relation:

$n_{g}\equiv c/v_{g}=cT_{1}/L=1+c\tau_{1}/\ell=1+c\rho\tau_{1}\sigma$. \qquad(2.13)

If there are the superluminal (instantaneous) jumps on distance $\Delta\ell$
at each scattering act, then $N\rightarrow N\prime=L/(\ell+\Delta\ell)$ and correspondingly

$n_{g}\rightarrow n_{g}\prime=1+c\tau_{1}/(\ell+\Delta\ell)=1+(n_{g}%
-1)(1+\Delta\ell/\ell)^{-1}$. \qquad(2.14)

In the region of anomalous dispersion, where $v_{g}>c$, this definition can
lead formally to $n_{g}<0$.

Far from all resonances the scattering of low energy photons goes as the
scattering on free electrons. Therefore, in accordance with the uncertainty
principle, the delay duration can be estimated as $\tau_{2}\approx1/2\omega$
and since the group and phase indices of refraction are close, then such
(rough) estimation for the free path length follows (2.13):

$\ell=c\tau_{1}/(n_{g}-1)\approx c/2\omega(n-1)$. \qquad(2.15)

Since the $s$-photon instantaneously jumps with each scattering act onto the
length $\Delta\ell$, the sum of pathes of this photon executing with the speed
$c$ would be equal to $L_{eff}=L-N\prime\Delta\ell$. Hence the ratio of
photons mean velocity in transparent media to the light speed in vacuum can be
estimated as

$u/c\simeq L/L_{eff}=1+\Delta\ell/\ell=1+2\pi(n-1)$. \qquad(2.16)

Just this expression must be comparing with experimental data. The most
representative experimental data are considered in [3] that had shown the
accordance of their results with the offered theory.

So for the light passage through the lightguide of sufficiently small radius,
when light waves almost on all their pass has an evanescent character, the
estimation (2.18) gives for typical $n=1.6$ the value: $u=4.77c$, which
corresponds to the experimental data [1, 3].

Note that this magnitude represents the maximal light speed in usually used
materials and therefore its value should have an exclusive significance for
optoelectronics, etc.

\section{TEMPORAL FUNCTIONS OF MASSIVE PARTICLES INTERACTION}

The temporal peculiarities of interactions of massive particles can be
considered in the close analogy with procedures examined above. But now we
must analyze the slightly more complicate Green functions $\Delta
_{c}(E,\mathbf{r})=\overline{\Delta}+i\Delta_{1}$, where $\Delta
_{1}(E,\mathbf{r})=(1/2r)\sin(E^{2}-m^{2})^{1/2}r$ (here and below in this
Section $\hbar=c=1$).

In the analogy with (2.6)

$\tau(E,\mathbf{r})=\frac{\partial}{i\partial E}\ln\Delta_{1}(E,\mathbf{r}%
)=\frac{irE}{(E^{2}\ -\ m^{2})^{1/2}}\cot(E^{2}-m^{2})^{1/2}r$.\qquad(3.1)

and the relations with $E > m$ and $E < m$ must be separately considered

If $E>m$, we have

$\tau_{1}=0$,

$\tau_{2}\rightarrow-\frac{E}{E^{2}\ -\ m^{2}}-2\sum\frac{1}{(E^{2}%
\ -\ m^{2})r^{2}\ -\ \pi^{2}n^{2}}$. \qquad(3.2)

As in the contrast to QED an undressed static state is here absent, the main
part of duration of particle (state) formation can be presented by the first term:

$\tau_{2}(E,\mathbf{r})\approx-E/(E^{2}-m^{2})\sim-1/2\Delta E$,\qquad\ (3.2')

which corresponds to the usually written form of energy-time uncertainty
relation, but with negative sign. It means that the considered transition must
be of advanced or instantaneous type, of tunneling character.

But if $E<m$, i.e. at consideration of coupled particles, then

$\tau_{1}=\frac{rE}{(E^{2}\ -\ m^{2})^{1/2}}\coth(E^{2}-m^{2})^{1/2}r$;
$\qquad\tau_{2}=0$.\qquad(3.3)

Therefore it formally seems that in this case the possibilities of
instantaneous transitions are absent. But if, as will be shown below, the
product $\tau(E,\mathbf{r})(E^{2}-m^{2})^{1/2}$ would be considered, it will
become evident that for such case $\tau_{1}$ and $\tau_{2}$ must be
interchanged, and the possibilities of instantaneous transitions still exist.

\section{INDETERMINATENESS AND UNCERTAINTY RELATIONS, EVANESCENT
PARTICLES}

The establishments of precise expressions for energy-time uncertainties for
different cases are crucial for our discussion. Therefore we must reconsider
the deduction of these relations with demonstration of their differences.

The most simple case of deduction is such one. The probability of transition
between states of one type with energies $E$ and $E\prime=E+\Delta E$ at
perturbation independent of time is proportional to the expression (e.g. [16]):

$W\sim\Delta E^{-2}\sin^{2}\Delta E\tau/2\hbar$. \qquad(4.1)

Its maximum is determining as $\Delta E\ \tau/2\hbar=\pi/2+n\pi$, where both
values can be negative in the complete correspondence with the condition
(1.1). But such approach is restrained by specific conditions of scattering
theory, and it could not be considered as the general rule for all interaction
processes, including particles decay and their transmutations.

Let's turn to the method offered by Mandelstam and Tamm [6]. They began with
comparison of two quantum expressions for Hermitian operators $\mathbf{A}$ and
$\mathbf{H}$, the Hamiltonian, with the standard deviations $\Delta A$ and
$\Delta H$:

$\Delta H\cdot\Delta A=\left\vert \left\langle \mathbf{HA}-\mathbf{AH}%
\right\rangle \right\vert $; \qquad(4.2)

$\hbar\partial_{t}\left\langle A\right\rangle =i\left\vert \left\langle
\mathbf{HA}-\mathbf{AH}\right\rangle \right\vert $, \qquad(4.2')

which leads together to the inequality

$\Delta H\cdot\Delta A\geq%
\frac12
\hbar\left\vert \partial_{t}\left\langle \mathbf{A}\right\rangle \right\vert
$. \qquad(4.3)

For its analysis the projector of some definite state $\psi_{0}$
must be introduced:

$\mathbf{P}=(\psi_{0},\psi)\psi_{0}$, \qquad$\mathbf{P}^{2}=\mathbf{P}$,
\qquad$\left\langle \mathbf{P}\right\rangle \leq1$, \qquad(4.4)

its standard deviation is defined as

$\Delta P=(\left\langle \mathbf{P}^{2}\right\rangle -\left\langle
\mathbf{P}\right\rangle ^{2})^{1/2}=(\left\langle \mathbf{P}\right\rangle
-\left\langle \mathbf{P}\right\rangle ^{2})^{1/2}$. \qquad(4.5)

The substitution of (4.5) in (4.3) brings the main Mandelstam-Tamm relation:

$\Delta H\ (\left\langle \mathbf{P}\right\rangle -\left\langle \mathbf{P}%
\right\rangle ^{2})^{1/2}$ $\geq%
\frac12
\hbar\left\vert \partial_{t}\left\langle \mathbf{A}\right\rangle \right\vert $
, \qquad(4.6)

satisfactory solution of which would be seeking in the form:

$\left\langle \mathbf{P}(t)\right\rangle \geq\cos^{2}(\alpha t)$. \qquad(4.7)

Here with some deflection from the initial consideration, related to the
lifetime of excited or intermediate state, we can propose that the completion
of transition means the return to the stable state after the definite time
duration, and therefore the substitution of $\left\langle \mathbf{P}%
(t)\right\rangle $ at $t=\tau$ in (4.6) leads to the condition (1.1):

$(\Delta H\ \tau)_{stable}=\pi\hbar$, \qquad(4.8)

just corresponding to our results in the Section 2.

If $(E^{2}-m^{2})<0$ and $\Delta H$ is pure imaginary, the magnitudes
$\tau_{1}$ and $\tau_{2}$ are interchanged; the solution of relation (4.6)
must be seeking in the form:

$\sinh^{2}(\alpha t)\leq\left\langle \mathbf{P}(t)\right\rangle \ \leq1$,
\qquad(4.9)

which culminates near to $\alpha\left\vert \tau\right\vert \sim1$. It means
the existence of uncertainty relation (1.3),

$(\Delta H\ \tau)_{virtual}\sim\frac{1}{2}\hbar$, \qquad(4.10)

which formally coincides with the usual forms for such relations. The
differences consist, of course, in possibilities of simultaneously pure
imaginary values of both standard deviations.

It must be noted that the original result of [6] remains completely valid for
the half time of decay state as (1.2).

Thus all examined types of processes must be considered separately. But as it
must be underlined, the initial expressions (4.2) show that both standard
deviations must have the same sign. Therefore it does not forbid the existence
of negative $\Delta\tau$\ at negative $\Delta H$ corresponding to the theorem,
cited above.

Note that the restrictions connected with the Hermitian character of
considered operators can be avoided by the most general formal deduction of
uncertainty relations given by Schr\"{o}dinger in [5]. The decomposition of
the arbitrary operators' product on the Hermitian and anti-Hermitian parts can
be taken as

$\mathbf{AB}=%
\frac12
(\mathbf{AB}+\mathbf{BA})+%
\frac12
(\mathbf{AB-BA})$. \qquad(4.11)

The subsequent quadrature of this expression, its averaging over complete
system of $\psi$-functions and replacement for operators on difference of
operators and their averaged values $\mathbf{A}\rightarrow\mathbf{A-}\Delta A$
, $\mathbf{B}\rightarrow\mathbf{B}-\Delta B$ bring to such expression for
standard deviations:

$(\Delta A)^{2}(\Delta B)^{2}\geq%
\frac14
\left\vert \left\langle \mathbf{AB-BA}\right\rangle \right\vert ^{2}+%
\frac14
[\left\langle \mathbf{AB}+\mathbf{BA}\right\rangle -2\left\langle
A\right\rangle \left\langle B\right\rangle ]^{2}$, \qquad(4.12)

which differs from the more usual form by the last term and can strengthen the
condition (4.2). The Heisenberg limit of this expression with $\left\vert
\left\langle \mathbf{AB-BA}\right\rangle \right\vert ^{2}\rightarrow\hbar^{2}$
and omitting of the second term shows a minimal value of uncertainties, which
can be achieved for pure states, in the weakly correlated conditions. But this
limit can be exceeded for some physical magnitudes (compare [4], such
possibilities are mentioned in other investigations also, e.g. [17]).

It must be noted that these relations can take place for the motion along one
of coordinates only. So Wigner [18] specially underlined that these
uncertainties depend on coordinates points, and if the process is progressing
in the $z$ direction:

$(\Delta t(z))^{2}=\int dxdydt(t-t_{0})^{2}\left\vert \psi(x,y,z,t)\right\vert
^{2}/\int dydt\left\vert \psi(x,y,z,t)\right\vert ^{2}$,

$(\Delta E(z))^{2}=\int dxdydE(E-E_{0})^{2}\left\vert \psi(x,y,z,E)\right\vert
^{2}/\int dydE\left\vert \psi(x,y,z,E)\right\vert ^{2}$,\qquad\ (4.13)

and they can be different, in general case, for other space axes.

This peculiarity can be the starting point at investigation of the phenomena
of frustrated total internal reflection (FTIR). In these phenomena the passing
photons can propagates with subluminal speed along the waveguide axis, but the
shifts in the perpendicular directions can be simultaneously instantaneous,
and it can lead to the suitable interpretation of proper observations.

\section{PARTICLE PHYSICS}

Let's consider possibility of particles instantaneous tunneling into the state
of higher mass; other characteristics can be transformed at such tunneling
also and therefore it will lead to particles transmutation. Such processes can
go with observance the energy-moment conservation, of course, i.e. at
interaction with matter and must correspond to the condition (4.10) (here and
below $c=1$):

$(\Delta m\ \Delta\tau)_{transmut}\sim\frac{1}{2}\hbar$, \qquad(5.1)

where $\Delta m=m_{i}-m_{f}<0$ and, correspondingly, $\Delta\tau_{transmut}%
<0$. The processes of such transitions are experimentally discovered yet for
neutral particles only. Therefore we begin with mesons and then will turn to
the more uncertain processes with neutrinos.

First candidates to such transition are $K^{0}$ mesons. As $m_{K_{L}}%
>m_{K_{S}}$, only the transition $K_{S}\rightarrow K_{L}$ via the considered
instantaneous transition seems possible. It is confirmed by the absence or by
extremely rarity of $K_{S}$ formation in the $K^{0}$ beam after definite
distance, i.e. by impossibility or by the weakness of superweak interaction,
usually attributable to the $CP$ nonconservation. But we can consider such
rare transitions as the opposite ones relative to the offered transmutations
into higher mass states. The estimation of relative magnitudes of direct and
opposite probabilities seems so far impossible.

Let's consider the experimental data for the "direct" process: $m_{K_{L}%
}-m_{K_{S}}=3.48\cdot10^{-12}$ MeV, $\left\vert \Delta\tau_{int}\right\vert
\rightarrow\tau_{K_{S}}=0.896\cdot10^{-10}$ s [19] leads to the estimation:

$\left\vert m_{K_{L}}-m_{K_{S}}\right\vert \ \tau_{K_{S}}=0.47\ \hbar$, \qquad(5.2)

i.e. they almost completely conform (5.1).

For $B^{0}$ mesons the mass difference $m_{B_{H}}-m_{B_{L}}=3.3\cdot10^{-10}$
MeV is known, but only the mean life time is determined: \ $\overline{\tau
}=1.55\cdot10^{-12}$ s. They give such estimation:

$\left\vert m_{B_{H}}-m_{B_{L}}\right\vert \ \overline{\tau}=0.775\ \hbar$,
\qquad(5.3)

but this value can be decreased if the life time of $B_{L}^{0}$ is lesser than
of its partner.

For $B_{s}^{0}$ is determined only that corresponding $\Delta m>94.8\cdot
10^{-10}$ MeV. It gives with (5.1) the estimation for the life time of
$L$-partner: $\tau_{L}<5.5\cdot10^{-14}$ s.

The known uncertainty of $D^{0}$ parameters does not allow such estimations.

There are many other pairs of mesons, including charged ones, with such mass
differences and times of decay, that can be considered as candidates into
analogical transmutation processes, but it requires a special considerations
far from our general aim. There is not seemingly any restrictions on
analogical permutations among fermions also.

For neutrinos it can be assumed that their masses, if they exist, correspond
to the commonly accepted hierarchy: $m_{\nu_{e}}<m_{\nu_{\mu}}<m_{\nu_{\tau}}$
or to the three mass eigenstates $m_{1}<m_{2}<m_{3}$ not associated with
particular lepton flavors (e.g. [20] and more recent reviews [21, 22]). Here
we do not distinguish these possibilities.

It means, in accordance with our general assumption, the possibilities of only
such instantaneous transmutations in matter to more massive partners:

$\nu_{e}\rightarrow\nu_{\mu}$; $\qquad\nu_{\mu}\rightarrow\nu_{\tau}$;
$\qquad\nu_{e}\rightarrow\nu_{\tau}$ \qquad(5.4)

and corresponding for antineutrino, but absence or suppression, at least, of
the opposite transitions.

Such discrimination evidently contradicts to the $CPT$ theorem. This uncleared
discrepancy can be connected at the instantaneous transition with the
violation of the Lorentz invariance needed for observance of this theorem
(this point evidently requires further investigations).

The arguments for this critical situation is related the absense of observed
transmutations of atmospheric $\nu_{\mu}$'s into $\nu_{e}$'s or sterile
states. Indeed, $\nu_{\mu}\rightarrow\nu_{e}$ transitions are also disfavored
by the Super-Kamiokande data, which prefer the $\nu_{\mu}\rightarrow\nu_{\tau
}$ channel [22].

So in accordance with our hypothesis there is not neutrinos oscillations, but
all their transmutations can be considered as the tunneling into states of
higher mass. (The hypotheses of neutrino oscillations, which have in mind as
the direct so the inverse transitions, were suggested independently in [23, 24].)

Note that the transmutations (5.4) with absence of opposite transitions can be
considered as the inevitable processes of specific ordering, i.e. as the
gradual phase transitions of neutrinos sets into possibly more passive states,
but with bigger masses.

Let's consider as an example of application of the proposed theory the
simplest and the most direct interpretation of the atmospheric neutrino
transition into the $\nu_{\mu}$ states [20]. It will give possibility for an
independent estimation of the neutrino mass.

The angular distribution of contained events shows that, for $E\sim1$ GeV, the
deficit of $\nu_{e}$'s comes mainly from $L\sim10^{2}\div10^{4}$ km. The
corresponding oscillation (transition) phase must be maximal:

$\Delta m^{2}($eV$^{2})L($km$)/2E($GeV$)\sim1$, \qquad(5.5)

which can be rewritten as

$\Delta m^{2}($eV$^{2})\tau($sec$)\sim\frac{2}{3}10^{-11}$.

The comparison with (5.1), rewritten as $\Delta m($eV$)\tau($sec$)\sim
~\frac{2}{3}10^{-15}$, leads to the estimation $\Delta m\sim10^{-4}$ eV, which
does not seems qualitatively inconsistent.

Note that this estimation determines the upper bound on the scale of
$\Lambda\sim10^{18}$ GeV, i.e. very close to the Planck scale $M_{Pl}%
\sim10^{19}$ GeV, and therefore it reduces the necessity for introduction a
new fundamental scale or so named New Physics.

\section{CONCLUSIONS}

Our examinations can be summed into such points.

1. The uncertainty or, more correctly, the indeterminateness principle must be
considered for each type of interaction separately, and although the
differences are only numerical, they can have different physical meaning.

2. Both standard deviations in the energy-time relations can be simultaneously
negatively or even pure imaginary ones. The negativity of the time standard
deviation can be interpreted as the instantaneous transition, the jump onto
definite distance or as the tunneling through classically forbidden region.

3. The visible nonlocality in the small, i.e. the violation of Lorentz
invariance can be considered on the same base as the violation of conservation
laws by virtual transitions, and it does not mean the violation of
relativistic causality by these evanescent particles, in the whole process.

4. The revealed forms of uncertainty relations demonstrate the peculiarities
of tunneling, and therefore they allow to explain and clarify the very old
problems of negative duration of tunneling at the standard quantum
calculations (e.g. [25]) as the natural feature of these transitions.

5. The new forms of uncertainties relations are deduced for transfer of some
excitations and for transmutation of particles into their more massive
partners. They allow some estimations and predictions of parameters of such
transitions or transmutations.

6. All these results prove that the possibility of some evanescent violations
of the classical Lorentz invariance are contained in the usual field theory
and the introducing of them ad hoc is not needed.

7. The examined processes can be considered as the violation of the
superselection rules ([26], in the modern form e.g. [27]), i.e. as the
transitions between different sectors of the Hilbert space through the
forbidden regions. They include the transitions between sectors with different
parities types, flavors and so on. Therefore it can be presupposed that these
transitions, completely or partly, are connected with violations of the
Lorentz invariance and take place simultaneously with it or owing to it.

These considerations allow proposing some attractive hypotheses for further examinations.

The most stimulating among them seems the possibility of intimately connection
of the Lorentz violation with violations of some discrete symmetries including
$CP$ or even $CPT$ and conservation laws.

Another interesting perspective consists in the suggestion of processes (5.4)
as the unique possible. They can lead, in particular, to the gradual
transmutation of all neutrinos into $\nu_{\tau}$'s as the final state of all
known neutrinos transmutations. If their mass is really bigger than of other
neutrinos, then just the tau-neutrinos can become the primary candidates into
the most part of a nonbarionic dark matter and essentially increase the
neutrino contribution into the critical density of the universe (compare [20]).

On the other hand such possibilities of the instantaneous transfer of
excitations in condensed states (macroscopic solids, atomic nucleus' and so
on) must be investigated, as they can fulfill everywhere a definite role in
the binding of constituents (cf. [28]).

\qquad

\section{ACKNOWLEDGEMENTS}

The author wishes to thank G. Nimtz and S. Elitzur for valuable
discussions, F.W. Hehl, I. I. Royzen, G. M. Rubinstein and M.
Tsindlekht for support and useful remarks in the course of the work.

\end{document}